\documentclass[a4paper,11pt]{article}
\usepackage{mathptm,times,graphicx}
\usepackage{amssymb,amsthm}
\usepackage{latexsym}
\usepackage{subfigure}

\newdimen\LENB \newdimen\LENW \newdimen\THI
\newdimen\LENWH \newdimen\LENTOT \newcount\N
\def\vbrknlnele#1#2#3{
 \LENB=#1pt \LENW=#2pt \THI=#3pt
 \LENWH=\LENW \divide\LENWH by 2
 \LENTOT=\LENB \advance\LENTOT by \LENW
 \vbox to \LENTOT{
   \vbox to \LENWH{}
   \nointerlineskip
   \vbox to \LENB{\hbox to \THI{\vrule width \THI height \LENB}}
   \nointerlineskip
   \vbox to \LENWH{}
 }}

\def\vbrknln#1{
 \N=#1
 \vcenter{
   \vbox{
     \loop\ifnum\N>0
       \vbox to 4pt{\vbrknlnele{2}{2}{0.1}}
       \nointerlineskip
       \advance\N by -1
     \repeat
 }}}

\def\hbrknlnele#1#2#3{
 \LENB=#1pt \LENW=#2pt \THI=#3pt
 \LENTOT=\LENB \advance\LENTOT by \LENW
 \vcenter{
   \vbox to \THI{
     \hbox to \LENTOT{
       \hfil
       \vrule width \LENB height \THI
       \hfil}
 }}}

\makeatletter

\usepackage{ulem}

\def\journal#1&#2,{\begingroup \let\journal=\dummyjournal
              \it #1\unskip~\bf\ignorespaces #2\rm,\endgroup}
\def\dummyjournal{\errmessage{Reference foul up: nested \journal macros}}

\def\eqref#1{(\ref{#1})}

\begin{document}
\title{
Note on the 2-component Analogue of
2-dimensional Long Wave-Short Wave Resonance Interaction System}
\author{
Ken-ichi Maruno\\
~{\small Department of Mathematics,
The University of Texas-Pan American,
Edinburg, TX 78541}\\
\quad \\
Yasuhiro Ohta\\
~{\small Department of Mathematics,
Kobe University,
Rokko, Kobe 657-8501, Japan}\\
\quad\\
Masayuki Oikawa\\
~{\small Research Institute for Applied Mechanics,
Kyushu University,
Kasuga, Fukuoka, 816-8580, Japan}
}
\date{\today}
\def\submitto#1{\vspace{28pt plus 10pt minus 18pt}
    \noindent{\small\rm To be submitted to : {\it #1}\par}}
\maketitle
\begin{abstract}
An integrable two-component analogue of the
two-dimensional long wave-short wave resonance
interaction (2c-2d-LSRI) system is studied.
Wronskian solutions of 2c-2d-LSRI system are presented.
A reduced case, which describes resonant interaction between an
interfacial wave and two surface wave packets in a two layer fluid, is also
 discussed.
\par
\kern\bigskipamount\noindent
2000 Mathematics Subject Classification. 35Q51, 35Q55, 37K40
\end{abstract}

\kern-\bigskipamount


\section{Introduction}
\quad \\
In these past decades, vector soliton equations have received so
much attention in mathematical physics and nonlinear physics
\cite{APT,manakov,laksh,APT2}.
Recently, we derived the following system in a two-layer fluid using
reductive perturbation method,
which was motivated by a paper by Onorato et. al.
\cite{yo-fluid,onorato}:
\begin{eqnarray}
&&{\rm i}(S^{(1)}_{t}+S^{(1)}_{y})-S^{(1)}_{xx}+LS^{(1)}=0\,,
\quad {\rm i}(S^{(2)}_{t}-S^{(2)}_{y})-S^{(2)}_{xx}+LS^{(2)}=0\,,\nonumber
\\
&&L_t=2(|S^{(1)}|^2+|S^{(2)}|^2)_x\,.
\label{2dls-fluid}
\end{eqnarray}
This system is an extension of the two-dimensional
long wave-short wave resonance
interaction system\cite{yajima-oikawa,oikawa}
and describes the two-dimensional resonant interaction between an
interfacial gravity wave and two surface gravity packets
propagating in directions symmetric about the propagation direction of the
interfacial wave in a two-layer fluid.

In this paper, we will study this system and its integrable modification,
\begin{eqnarray}
&&{\rm i}(S^{(1)}_{t}+S^{(1)}_{y})-S^{(1)}_{xx}+LS^{(1)}
=2{\rm i}{S^{(2)}}^*Q\,,\,\nonumber\\ 
&&{\rm i}(S^{(2)}_{t}-S^{(2)}_{y})-S^{(2)}_{xx}+LS^{(2)}=2{\rm
i}{S^{(1)}}^*Q\,,
\nonumber
\\
&& L_t=2(|S^{(1)}|^2+|S^{(2)}|^2)_x\,,\qquad 
Q_x=S^{(1)}S^{(2)}\,. \label{2dls-int}
\end{eqnarray}
where ${}^*$ means complex conjugate.
In our recent paper \cite{omo07}, we studied
\begin{eqnarray}
&&{\rm i}(S^{(1)}_{t}+S^{(1)}_{y})-S^{(1)}_{xx}+LS^{(1)}=0\,,
\quad {\rm i}(S^{(2)}_{t}+S^{(2)}_{y})-S^{(2)}_{xx}+LS^{(2)}=0\,,
\nonumber
\\
&&L_t=2(|S^{(1)}|^2+|S^{(2)}|^2)_x\,.
\end{eqnarray}
Note that this system is different from the system (\ref{2dls-fluid})
only in the sign of $y$-derivative term $S_y^{(2)}$.

\section{Bilinear Forms and Wronskian Solutions}

Consider a two-component analogue of two-dimensional
long wave-short wave resonance interaction (2c-2d-LSRI) system
(\ref{2dls-int}).
Using the dependent variable transformation 
$L=-(2\log F)_{xx},\,
S^{(1)}=G/F,\,
S^{(2)}=H/F,\,
Q=-K^*/F\,,
$
we obtain
\begin{equation}
\begin{array}{l}
(D_x^2-{\rm i}(D_t+D_y))G\cdot F=2{\rm i}H^*K^*\,,
\qquad D_xD_tF\cdot F=-2(GG^*+HH^*)\,,
\\
(D_x^2-{\rm i}(D_t-D_y))H\cdot F=2{\rm i}G^*K^*\,,
\qquad D_xK\cdot F=-G^*H^*\,.
\end{array}\label{bilin}
\end{equation}
These bilinear forms have the three-component Wronskian solution
\cite{DJKM1,DJKM2,HirotaBook}.

Consider the following three-component Wronskian:
\begin{eqnarray*}
&&\hskip-5mm\tau_{NML}=
\left|\,{\bf \varphi} \quad {\bf \psi} \quad {\bf \chi}\, \right|\,,
\end{eqnarray*}
where ${\bf \varphi}$, ${\bf \psi}$ and ${\bf \chi}$ are
$(N+M+L)\times N$, $(N+M+L)\times M$ and $(N+M+L)\times L$ matrices,
respectively:
${\bf \varphi}
=(\partial_{x_1}^{j-1}\varphi_i)_{1\le i\le N+M+L}^{1\le j\le N}$,
${\bf \psi}=(\partial_{x_1}^{j-1}\psi_i)_{1\le i\le N+M+L}^{1\le j\le M}$
and
${\bf \chi}=(\partial_{x_1}^{j-1}\chi_i)_{1\le i\le N+M+L}^{1\le j\le L}$,
and $\varphi_i$ is an arbitrary function of $x_1$ and $x_2$ satisfying
$\partial_{x_2}\varphi_i=\partial_{x_1}^2\varphi_i$,
and $\psi_i$ and $\chi_i$ are arbitrary functions of $y_1$ and $z_1$,
respectively.
The above Wronskian satisfies
\begin{eqnarray*}
&&(D_{x_1}^2-D_{x_2})\tau_{N+1,M-1,L}\cdot\tau_{NML}=0\,,
\qquad (D_{x_1}^2-D_{x_2})\tau_{N+1,M,L-1}\cdot\tau_{NML}=0\,,
\\
&&D_{x_1}D_{y_1}\tau_{NML}\cdot\tau_{NML}=2\tau_{N+1,M-1,L}\tau_{N-1,M+1,L}\,,
\\
&&D_{x_1}D_{z_1}\tau_{NML}\cdot\tau_{NML}=2\tau_{N+1,M,L-1}\tau_{N-1,M,L+1}\,,
\\
&&D_{x_1}\tau_{N,M+1,L-1}\cdot\tau_{NML}=-\tau_{N-1,M+1,L}\tau_{N+1,M,L-1}\,,
\\
&&D_{y_1}\tau_{N-1,M,L+1}\cdot\tau_{NML}=-\tau_{N,M-1,L+1}\tau_{N-1,M+1,L}\,,
\\
&&D_{z_1}\tau_{N+1,M-1,L}\cdot\tau_{NML}=-\tau_{N+1,M,L-1}\tau_{N,M-1,L+1}\,.
\end{eqnarray*}
Setting
\begin{eqnarray*}
f=\tau_{NML}\,,
&&g=\tau_{N+1,M-1,L}\,,
\qquad h=\tau_{N-1,M,L+1}\,,
\qquad k=\tau_{N,M+1,L-1}\,,
\\
&&\bar g=\tau_{N-1,M+1,L}\,,
\qquad \bar h=\tau_{N+1,M,L-1}\,,
\qquad \bar k=\tau_{N,M-1,L+1}\,,
\end{eqnarray*}
we have the following bilinear forms:
\begin{eqnarray*}
&&(D_{x_1}^2-D_{x_2})g\cdot f=0\,,
\qquad (D_{x_1}^2+D_{x_2})\bar g\cdot f=0\,,
\qquad D_{x_1}D_{y_1}f\cdot f=2g\bar g\,,
\\
&&(D_{x_1}^2+D_{x_2})h\cdot f=0\,,
\qquad (D_{x_1}^2-D_{x_2})\bar h\cdot f=0\,,
\qquad D_{x_1}D_{z_1}f\cdot f=2h\bar h\,,
\\
&&D_{x_1}k\cdot f=-\bar g\bar h\,,
\qquad\qquad D_{y_1}h\cdot f=-\bar g\bar k\,,
\qquad\qquad D_{z_1}g\cdot f=-\bar h\bar k\,,
\\
&&D_{x_1}\bar k\cdot f=gh\,,
\qquad\qquad\quad D_{y_1}\bar h\cdot f=gk\,,
\qquad\qquad\ \ D_{z_1}\bar g\cdot f=hk\,.
\end{eqnarray*}
By the change of independent variables
$x_1=x\,,\,
x_2=-{\rm i}y\,,
\,
y_1=y-t\,,
\,
z_1=-y-t\,
\, (x,y,t:\hbox{real}),
$
we have
$
\partial_x=\partial_{x_1}\,,
\,
\partial_y=-{\rm i}\partial_{x_2}+\partial_{y_1}-\partial_{z_1}\,,
\,
\partial_t=-\partial_{y_1}-\partial_{z_1}\,.
$
Thus we obtain
\begin{eqnarray*}
&&(D_x^2-{\rm i}(D_t+D_y))g\cdot f=-2{\rm i}\bar h\bar k\,,
\qquad (D_x^2+{\rm i}(D_t+D_y))\bar g\cdot f=-2{\rm i}hk\,,\\
&&(D_x^2-{\rm i}(D_t-D_y))h\cdot f=-2{\rm i}\bar g\bar k\,,
\qquad (D_x^2+{\rm i}(D_t-D_y))\bar h\cdot f=-2igk\,,
\\
&& D_xD_tf\cdot f=-2(g\bar g+h\bar h)\,,
\qquad D_xk\cdot f=-\bar g\bar h\,,
\qquad D_x\bar k\cdot f=gh\,.
\end{eqnarray*}
Consider solutions satisfying the following condition
\begin{equation}
\bar g{\cal G}=(g{\cal G})^*\,,
\qquad
\bar h{\cal G}=(h{\cal G})^*\,,
\qquad
\bar k{\cal G}=-(k{\cal G})^*\,,
\qquad f{\cal G}:\hbox{real}\,,\label{condition}
\end{equation}
where ${\cal G}$ is a gauge factor.
Then, for
$F=f{\cal G}$, $G=g{\cal G}$, $H=h{\cal G}$, $K=k{\cal G}$,
we will obtain
the bilinear equations of the 2c-2d-LSRI system (\ref{bilin}).
Thus the 2c-2d-LSRI system has a three-component Wronskian solution.

To satisfy the condition (\ref{condition}), we consider the
following constrained case:
$N=M+L$, $\psi_i=0$ for $2M+1\le i\le 2M+2L$, $\chi_i=0$ for $1\le i\le 2M$
and
\begin{eqnarray*}
&&\varphi_i=e^{\xi_i}\,,
\qquad
\varphi_{M+i}=e^{-\xi_i^*}\,,\quad
\xi_i=p_ix_1+p_i^2x_2\,,
\\
&&\psi_{i}=a_ie^{\eta_i}\,,
\qquad
\psi_{M+i}=a_{M+i}e^{-\eta_i^*}\,,\qquad
\eta_i=q_iy_1+\eta_{i0}\,,
\end{eqnarray*}
for $i=1,2,\cdots,M$, and
\begin{eqnarray*}
&&\varphi_{2M+i}=e^{\theta_{i}}\,,
\qquad
\varphi_{2M+L+i}=e^{-\theta_i^*}\,,\qquad
\theta_i=s_ix_1+s_i^2x_2\,,
\\
&&\chi_{2M+i}=b_ie^{\zeta_i}\,,
\qquad
\chi_{2M+L+i}=b_{L+i}e^{-\zeta_i^*}\,,\qquad
\zeta_i=r_iz_1+\zeta_{i0}\,,
\end{eqnarray*}
for $i=1,2,\cdots,L$, where $p_i$, $s_i$, $q_i$, $r_i$ are wave numbers
and $\eta_{i0}$, $\zeta_{i0}$ are phase constants.
The parameters
$a_i$ and $b_i$ must be determined from the condition of complex conjugacy.
By using the standard technique \cite{HH}, $a_i$ and $b_i$ are determined as
\begin{eqnarray*}
&&a_i=\prod_{k=1\atop k\ne i}^M\frac{p_k-p_i}{q_k-q_i}
\prod_{k=1}^M\frac{p_k^*+p_i}{q_k^*+q_i},\quad
a_{M+i}=\prod_{k=1}^L(s_k+p_i^*)(s_k^*-p_i^*),\quad
1\le i\le M,
\\
&&b_i=\prod_{k=1\atop k\ne i}^L\frac{s_k-s_i}{r_k-r_i}
\prod_{k=1}^L\frac{s_k^*+s_i}{r_k^*+r_i},\quad
b_{L+i}=\prod_{k=1}^M(p_k+s_i^*)(p_k^*-s_i^*),\quad
1\le i\le L,
\end{eqnarray*}
and the condition (\ref{condition}) is satisfied for the gauge factor,
\begin{eqnarray*}
&&{\cal G}=\prod_{1\le i<j\le M}(p_j^*-p_i^*)(q_i-q_j)
\prod_{1\le i<j\le L}(s_j^*-s_i^*)(r_i-r_j)
\prod_{i=1}^M\prod_{j=1}^L(p_i-s_j)
\\
&&\quad\times
e^{\sum_{i=1}^M(\xi_i^*-\eta_i)+\sum_{j=1}^L(\theta_j^*-\zeta_j)}.
\end{eqnarray*}
This solution represents the $(M+L)$-soliton, i.e., $M$ solitons
propagate on the first component of short wave $S^{(1)}$ whose
complex wave numbers are given by $p_i$, $q_i$ and complex phase
constants are $\eta_{i0}$, and $L$ solitons propagate on the second
one $S^{(2)}$ whose complex wave numbers and phase constants are
$s_i$, $r_i$ and $\zeta_{i0}$.

\begin{figure}[t!]
\centerline{
\includegraphics[scale=0.35]{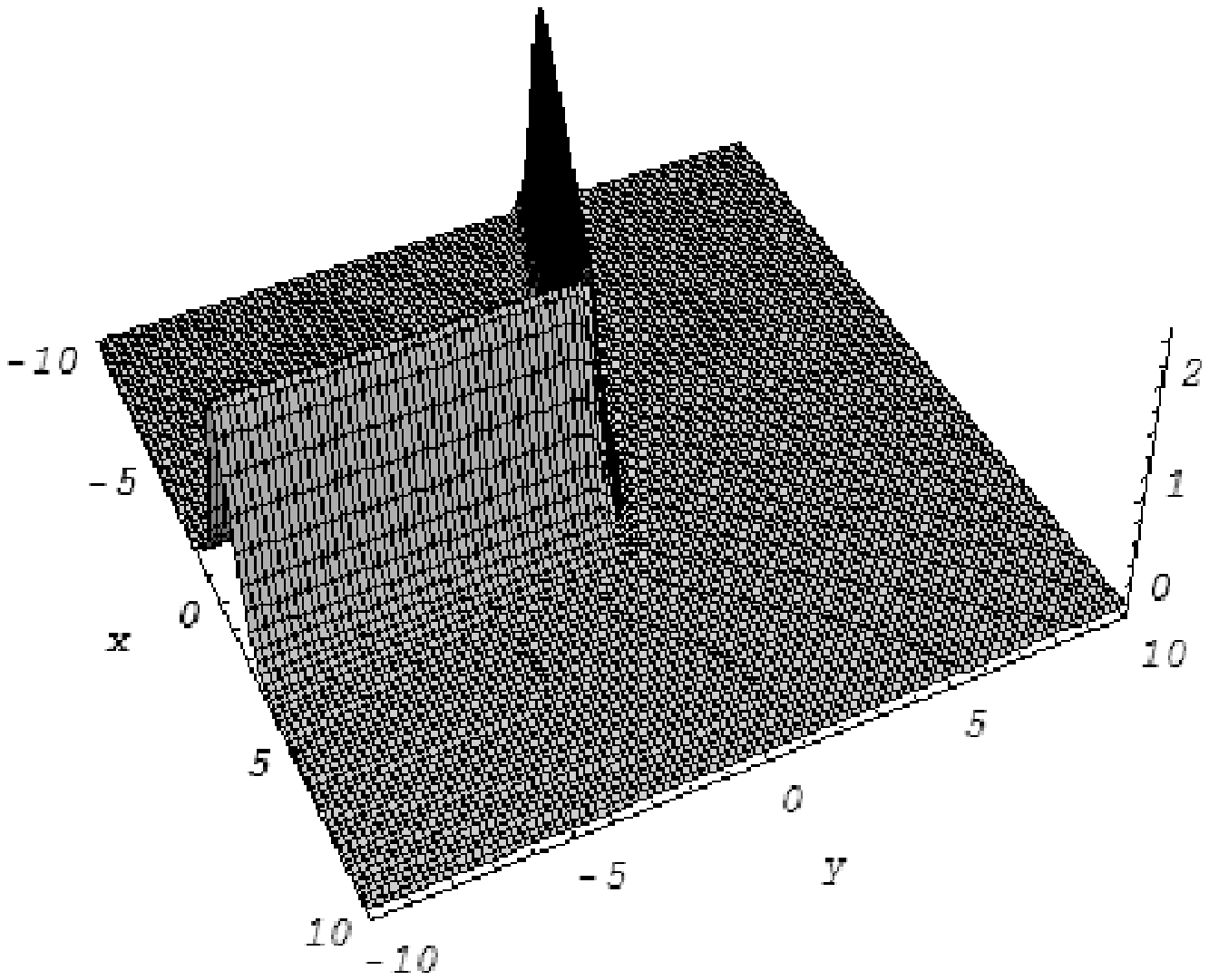}\quad
\includegraphics[scale=0.35]{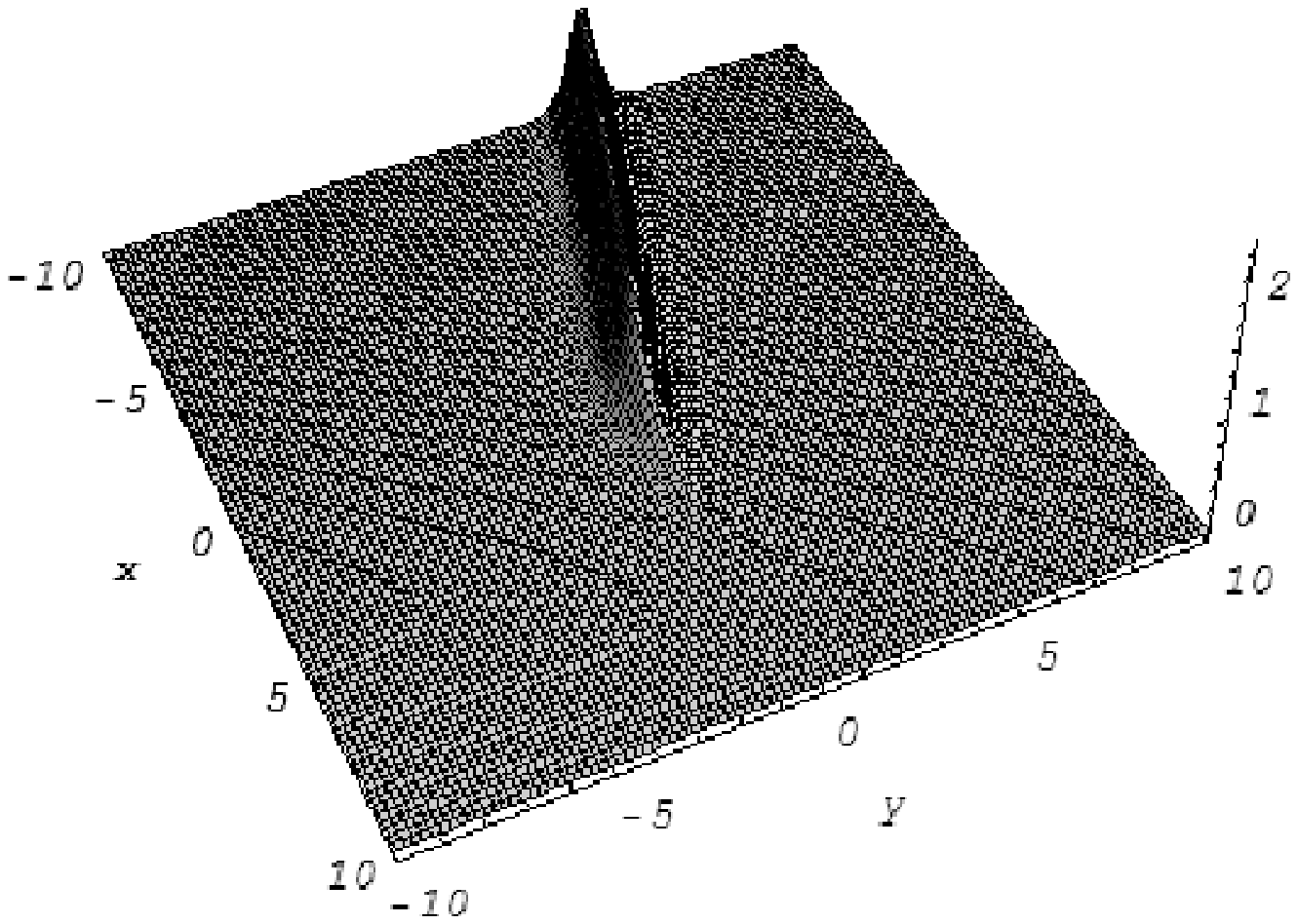}\quad
\includegraphics[scale=0.35]{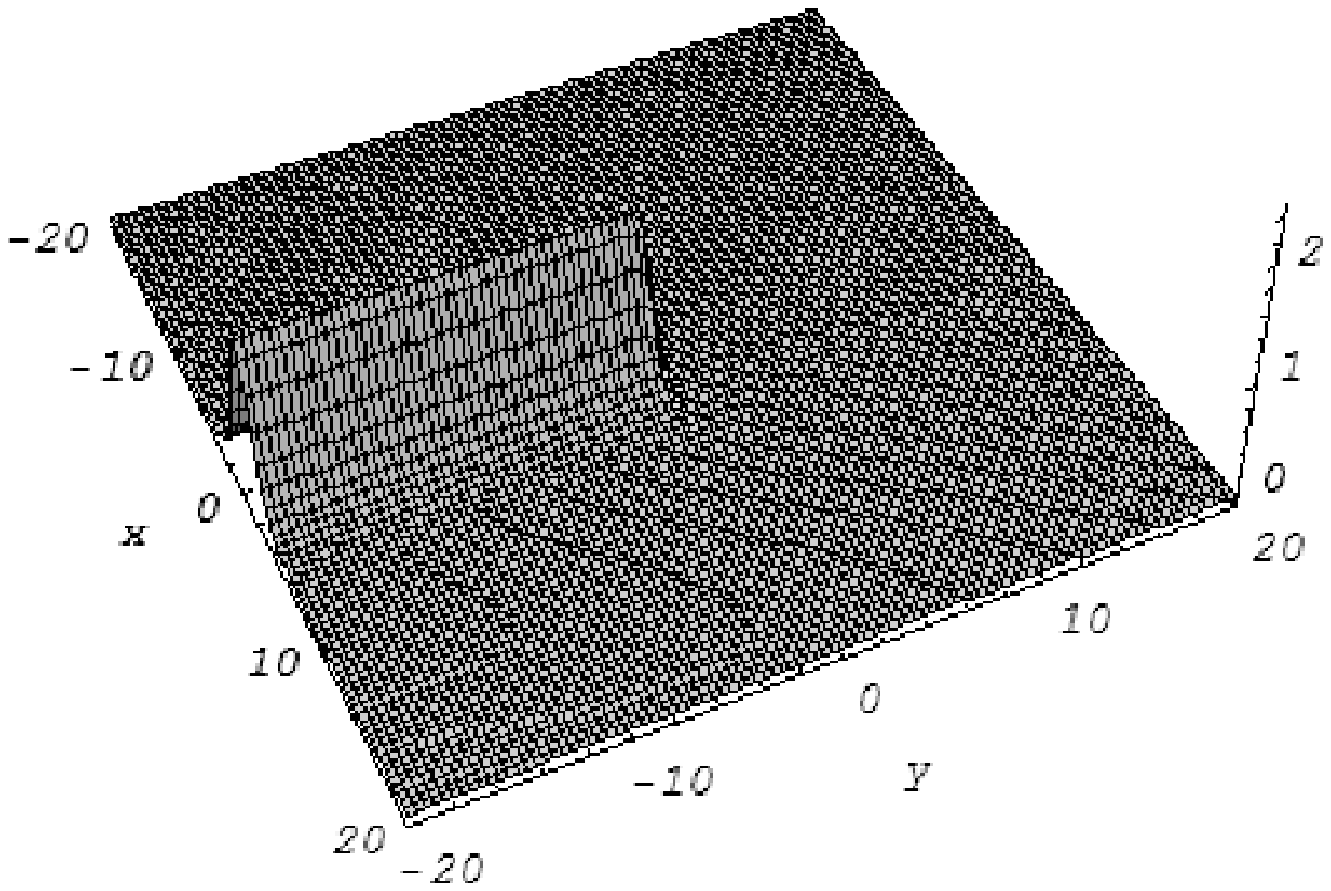}
}
\kern-0.3 \textwidth
\hbox to \textwidth{\kern -2em (a)\kern 5em \hss (b)\kern13em (c) \kern 7em}
\kern+0.355\textwidth
\centerline{
\includegraphics[scale=0.35]{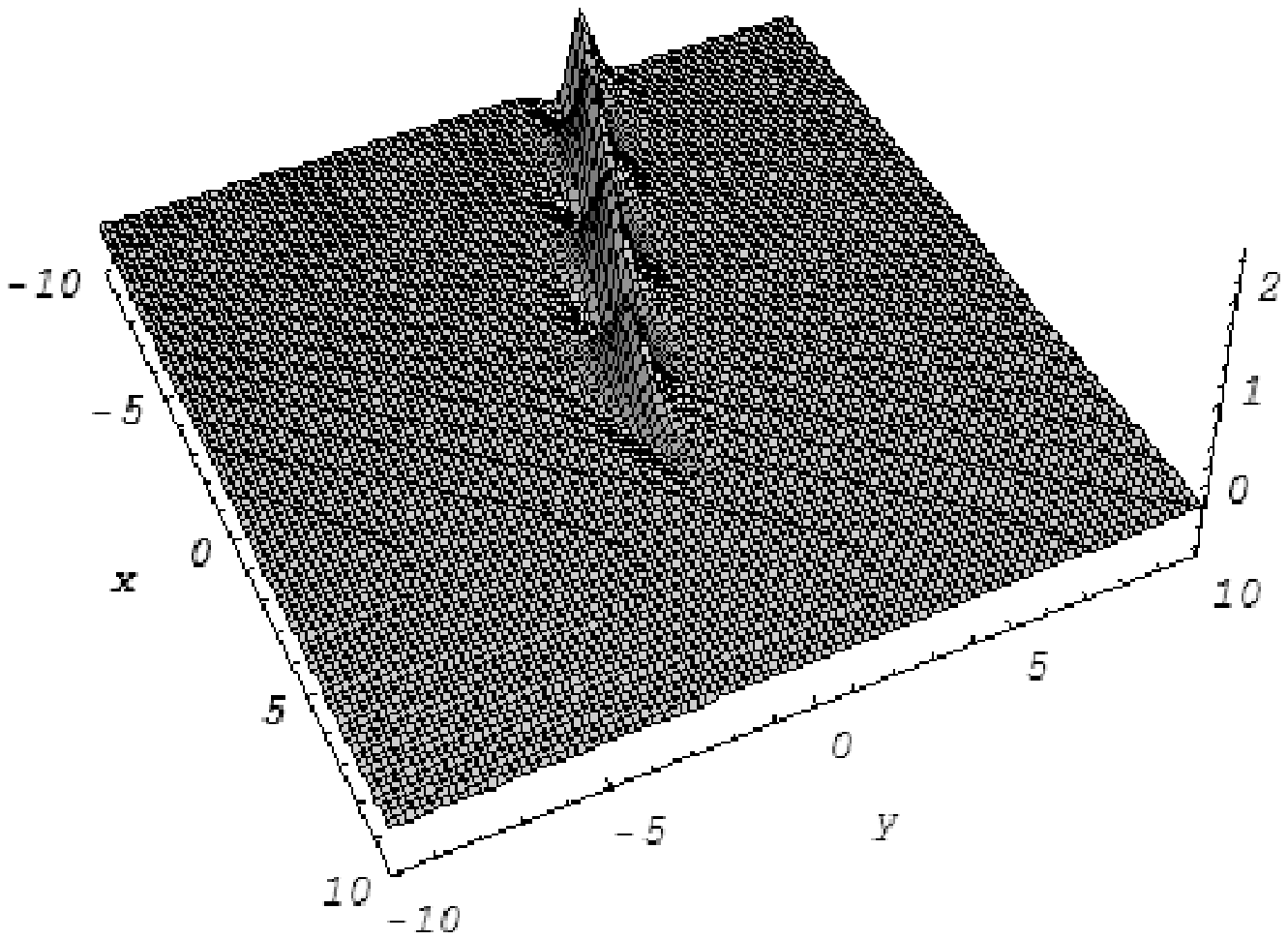}\quad
\includegraphics[scale=0.35]{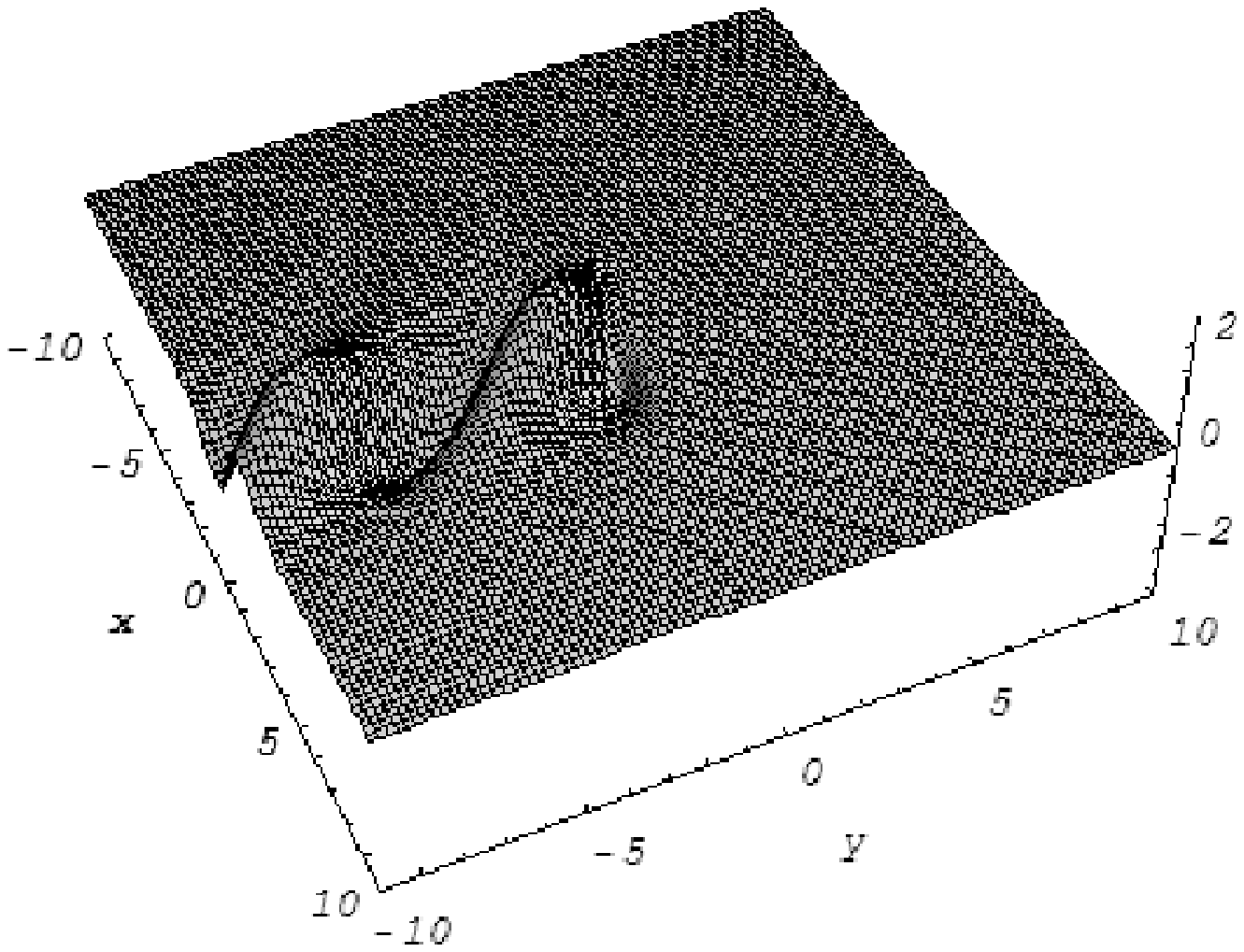}}
\kern-0.3\textwidth
\hbox to \textwidth{\hss(d)\kern10em\hss(e)\kern13em}
\kern+0.355\textwidth
\caption{Single line soliton of eqs.(\ref{2dls-int}), which 
is obtained by tau-functions of 
(\ref{singlesoliton}).
(a) $-L$\,, (b) $|S^{(1)}|$, (c) $|S^{(2)}|$, (d) ${\rm Re}\, [S^{(1)}]$,
(e) ${\rm Re}\, [S^{(2)}]$. 
The parameters are $p=1+{\rm i}, q=-1+2{\rm i}, r=-2+{\rm i}$.
}
\label{solitonsolution-int}
\end{figure}

For instance by taking $M=L=1$, (1+1)-soliton solution is given as
\begin{eqnarray*}
&&{\cal G}f=c\Big(\frac{p+p^*}{q+q^*}\frac{s+s^*}{r+r^*}\frac{1}{|p+s^*|^2}
-\frac{s+s^*}{r+r^*}e^{\xi+\xi^*-\eta-\eta^*}
-\frac{p+p^*}{q+q^*}e^{\theta+\theta^*-\zeta-\zeta^*}
\\
&&\qquad
+|p-s|^2e^{\xi+\xi^*-\eta-\eta^*+\theta+\theta^*-\zeta-\zeta^*}\Big),
\\
&&{\cal G}g=c(p+p^*)e^{\xi-\eta}\Big(\frac{s+s^*}{r+r^*}\frac{1}{p^*+s}
-(p-s)e^{\theta+\theta^*-\zeta-\zeta^*}\Big),
\\
&&{\cal G}h=-c(s+s^*)e^{\theta^*-\zeta^*}\Big(\frac{p+p^*}{q+q^*}
\frac{1}{p^*+s}+(p^*-s^*)e^{\xi+\xi^*-\eta-\eta^*}\Big),
\\
&&{\cal G}k=c\frac{(p+p^*)(s+s^*)}{p+s^*}e^{\xi^*-\eta^*+\theta-\zeta},
\end{eqnarray*}
where $c=-|(p-s)(p+s^*)|^2$ and we dropped the index 1 for simplicity.
In order to satisfy the regularity condition $f\ne0$, we can take
${\rm Re}\,p>0$, ${\rm Re}\,s>0$, ${\rm Re}\,q<0$ and ${\rm Re}\,r<0$.
After removing the gauge and constant factors,
by choosing the same wave number in $x$ direction for the above two solitons,
i.e., $s=p$, we obtain the single soliton solution,
\begin{eqnarray}
&&f=\frac{1}{p+p^*}-e^{\xi+\xi^*}
((q+q^*)e^{-\eta-\eta^*}+(r+r^*)e^{-\zeta-\zeta^*}),\nonumber
\\
&&g=(q+q^*)e^{\xi-\eta},\, h=-(r+r^*)e^{\xi^*-\zeta^*},\,
k=(q+q^*)(r+r^*)e^{\xi+\xi^*-\eta^*-\zeta},\label{singlesoliton}
\end{eqnarray}
where $\xi=px-ip^2y$, $\eta=q(y-t)+\eta_0$ and $\zeta=-r(y+t)+\zeta_0$. 
Figure \ref{solitonsolution-int} shows the plots of 
this single soliton solution. $L$ shows V-shape soliton, 
$|S^{(1)}|$ and $|S^{(2)}$ shows solitoff behaviour \cite{Gilson}.   

\section{Solutions in the case without $Q$}

We consider the 2c-2d-LSRI system (\ref{2dls-fluid}) without the fourth
field $Q$ in (\ref{2dls-int}).
This system (\ref{2dls-fluid}) describes waves in the two-layer fluid.
Setting
$
L=-(2\log F)_{xx},\,
S^{(1)}=G/F,\,
S^{(2)}=H/F\,,
$
we have
\begin{eqnarray*}
&&[{\rm i}(D_t+D_y)-D_x^2]G \cdot F=0\,,
\qquad [{\rm i}(D_t-D_y)-D_x^2]H \cdot F=0\,,\\
&&-(D_tD_x-2c)F\cdot F=2GG^*+2HH^*\,.
\end{eqnarray*}
Here we consider the case of $c=0$.

Using the procedure of the Hirota bilinear method, we
obtain the single soliton solution
\[
F=1+A_{11}\exp(\eta_1+\eta_1^*)\,,\quad G=a_1 \exp(\eta_1)  \,,\quad
H=b_1 \exp (\xi_1)\,,
\]
\[\eta_j=p_jx+{\rm i} q_jy+\lambda_j t+\eta_j^{(0)}\,,\qquad
\xi_j=p_jx-{\rm i} q_jy+\lambda_j t+\eta_j^{(0)}\,,\qquad
\]
\[
 A_{11}=-\frac{a_1a_1^*+b_1 b_1^*}
{(p_1+p_1^*)(\lambda_1+\lambda_1^*)}\,,\quad \lambda_1=-{\rm i}p_1^2-{\rm i}q_1\,.
\]
Here $q_j$ is a real number.
We can rewrite $A_{11}$ as
\[
 A_{11}=-\frac{a_1a_1^*+b_1 b_1^*}
{(p_1+p_1^*)^2({\rm i}p_1^*-{\rm i}p_1)}\,.
\]
Thus we have
\begin{eqnarray*}
&&S^{(1)}=\frac{a_1 \exp(\eta_1) }{
1+A_{11}\exp(\eta_1+\eta_1^*)
}
\,,\quad
S^{(2)}=\frac{b_1 \exp (\xi_1)}
{1+A_{11}\exp(\eta_1+\eta_1^*)}
\,,
\quad\\
&&L=-2\frac{\partial^2}{\partial x^2}\log (1+A_{11}\exp(\eta_1+\eta_1^*))\,.
\end{eqnarray*}
Since
$|S^{(1)}|^2=GG^*/F^2$, $|S^{(2)}|^2=HH^*/F^2$,
$L=-2\frac{\partial^2}{\partial x^2}\log F$
do not include $y$, all solitons
propagate in the $x$ direction.

There is an exact solution depending on $y$-variable,
\begin{eqnarray*}
&&S^{(1)}=\frac{A_1\exp(p x+q y+r t)}{
1+\exp(2(px+qy+rt))
}\exp({\rm i}(k_1x+l_1y+m_1t))\,,
\quad \\
&&S^{(2)}=\frac{A_2 \exp(px+qy+rt)}
{1+\exp(2(px+qy+rt))}
\exp({\rm i}(k_2x+l_2y+m_2t))
\,,
\quad\\
&&L=\frac{A\exp(2(px+qy+rt))}{(1+\exp(2(px+qy+rt)))^2}\,,
\end{eqnarray*}
where
$p$, $q$, $r$, $k_1$, $l_1$, $m_1$, $k_2$, $l_2$, $m_2$, $A_1$, $A_2$, $A$
satisfy the relations
$r=(k_1+k_2)p,\,
q=(k_1-k_2)p,\,
m_1=k_{1}^2-l_1-p^2,\,
m_2=k_{2}^2+l_2-p^2,\,
A=-8p^2,\, A_{1}^2+A_{2}^2=-4(k_1+k_2)p^2$,
and $p,q,k_1,l_1,l_2$ are arbitrary parameters.
\begin{figure}[t!]
\centerline{
\includegraphics[scale=0.35]{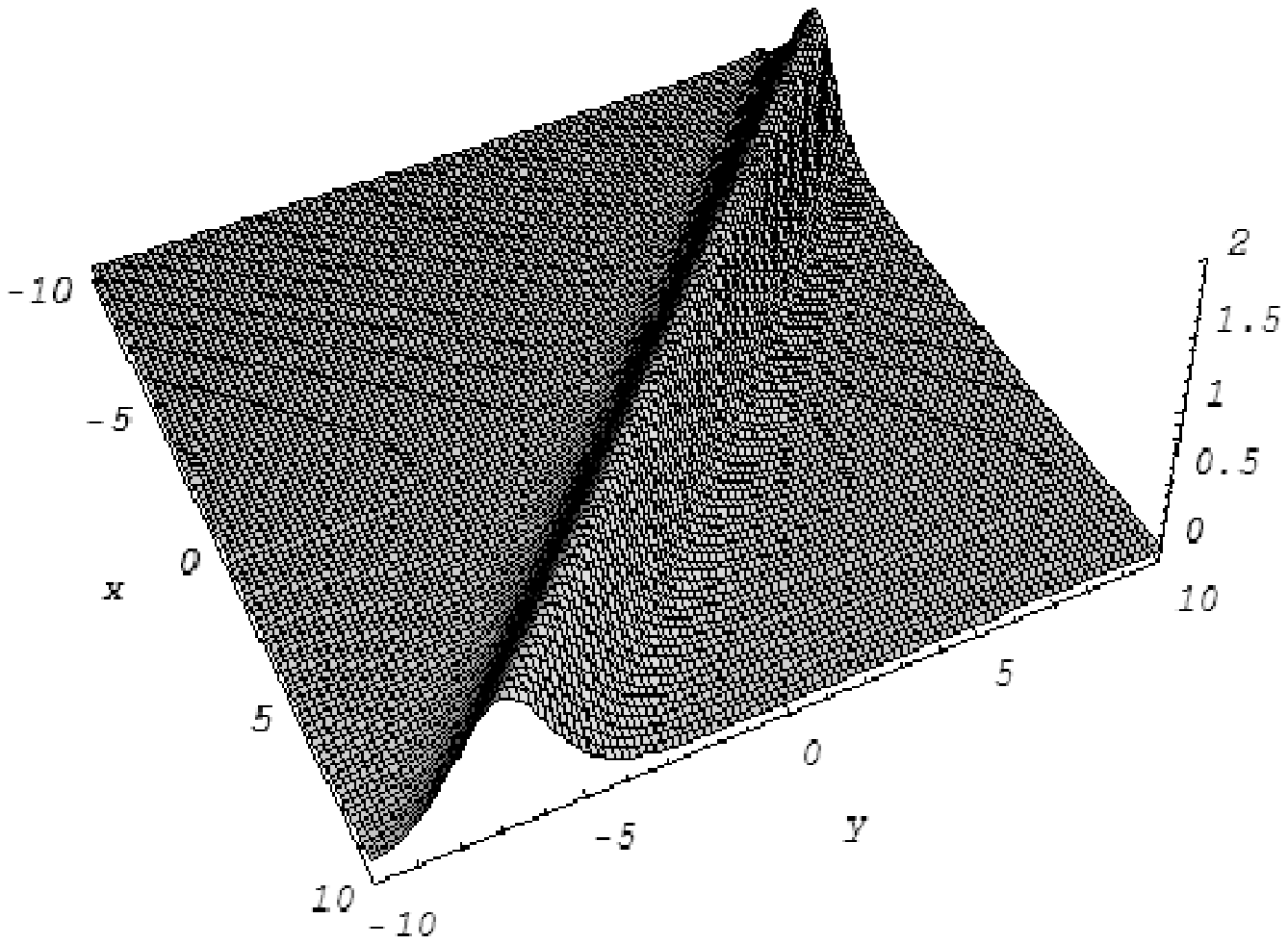}\quad
\includegraphics[scale=0.35]{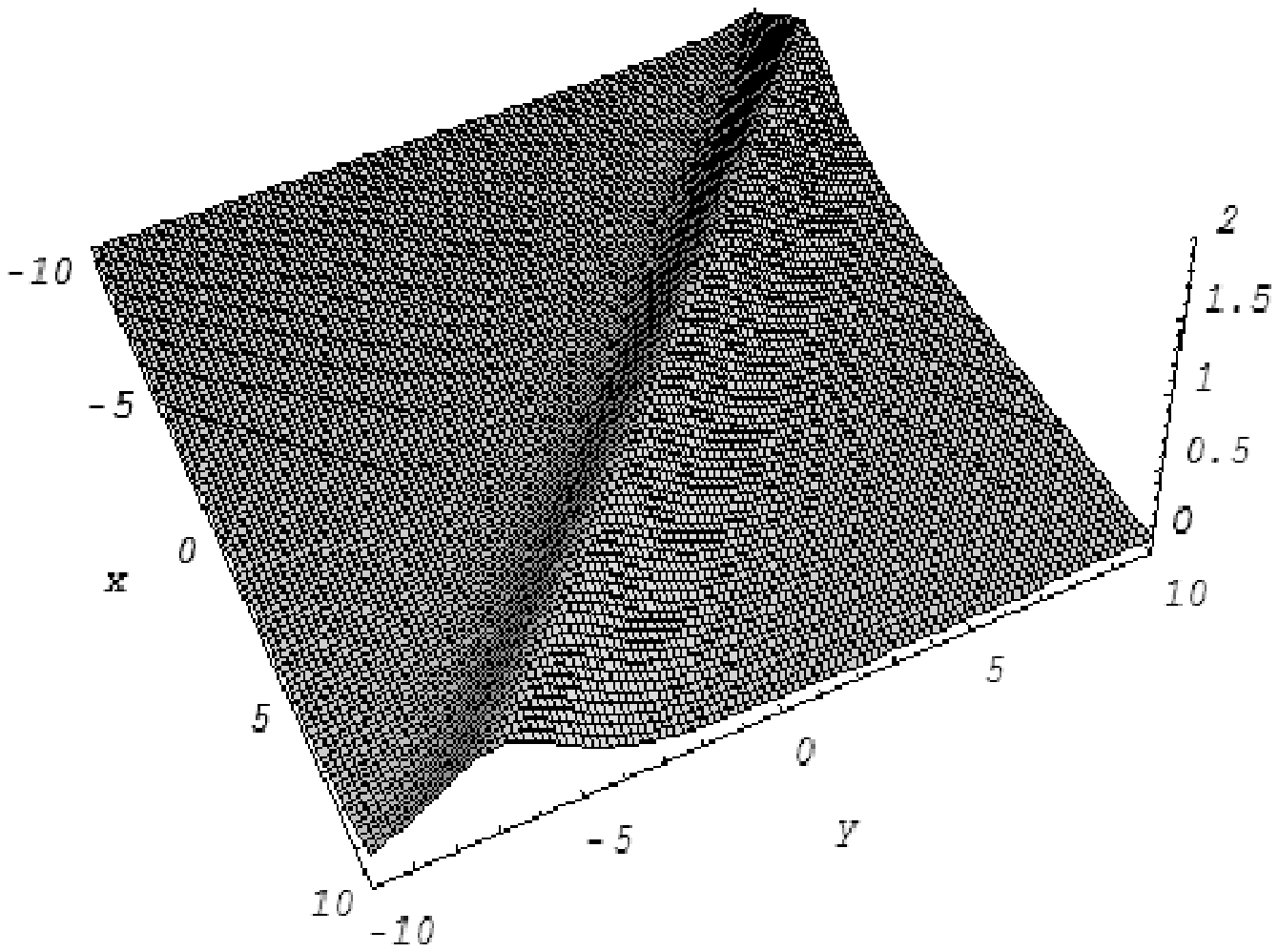}\quad
\includegraphics[scale=0.35]{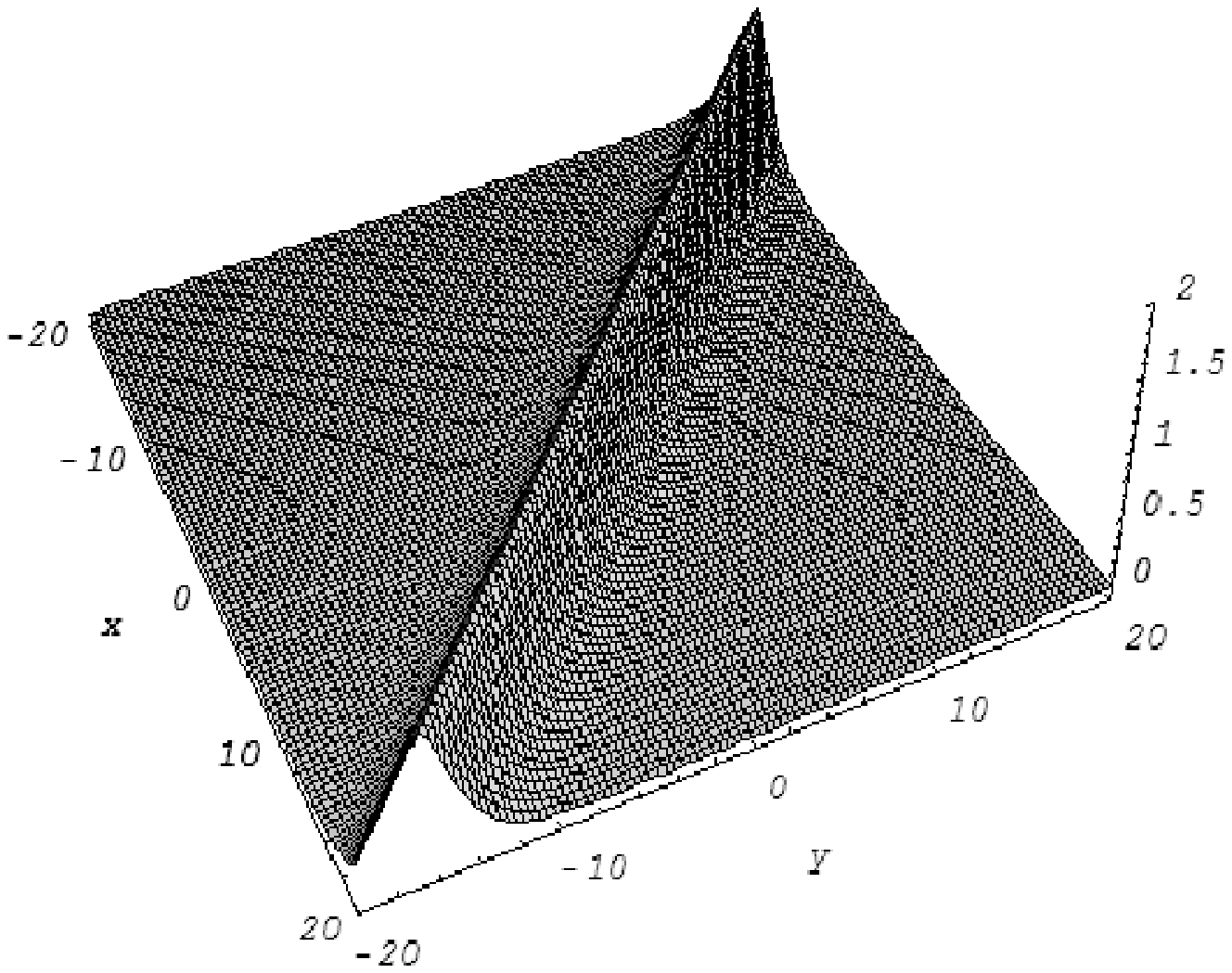}
}
\kern-0.3 \textwidth
\hbox to \textwidth{\kern -2em (a)\kern 5em \hss (b)\kern13em (c) \kern 7em}
\kern+0.355\textwidth
\centerline{
\includegraphics[scale=0.35]{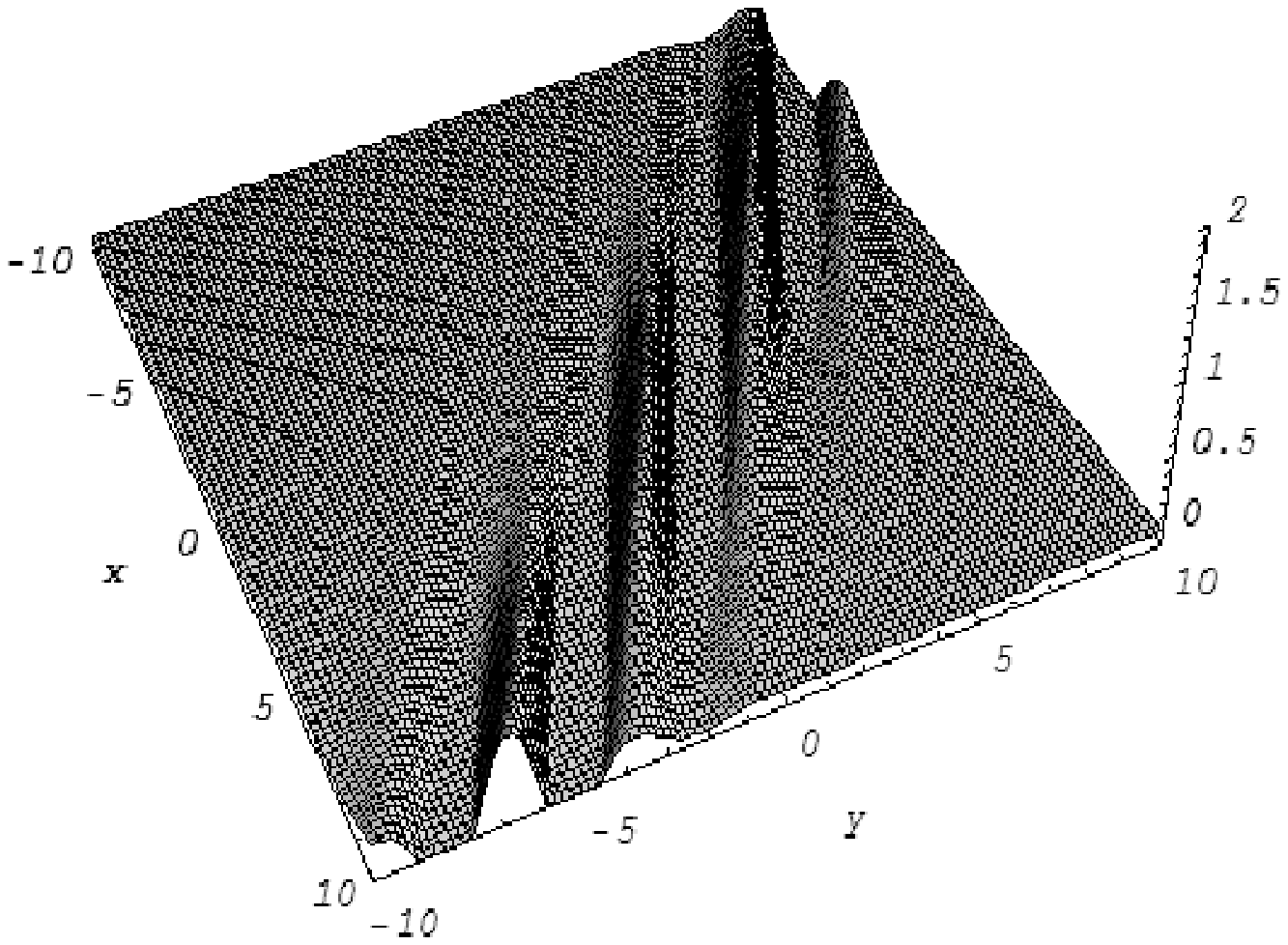}\quad
\includegraphics[scale=0.35]{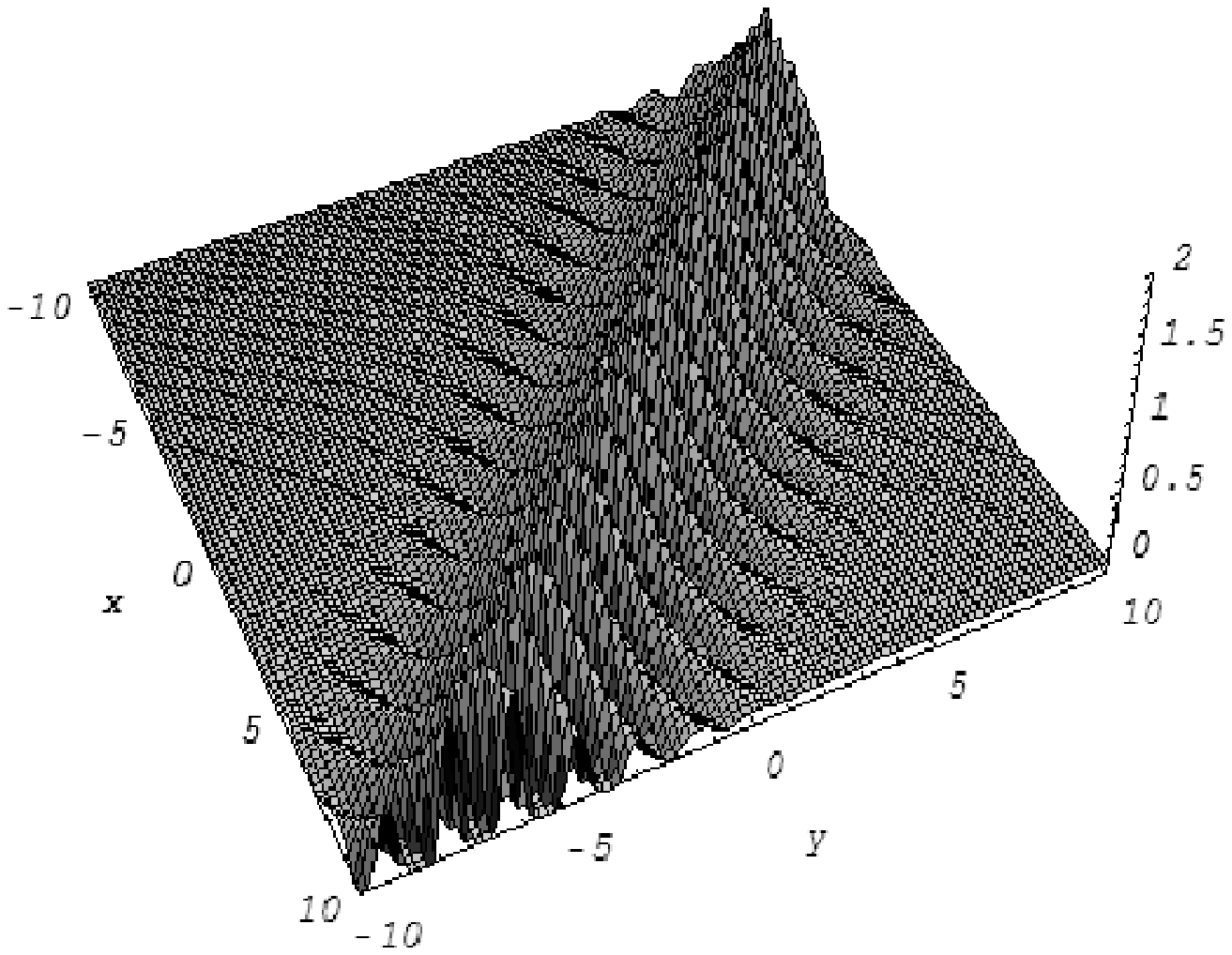}}
\kern-0.3\textwidth
\hbox to \textwidth{\hss(d)\kern10em\hss(e)\kern13em}
\kern+0.355\textwidth
\caption{Line soliton of eqs.(\ref{2dls-fluid}).
(a) $-L$\,, (b) $|S^{(1)}|$, (c) $|S^{(2)}|$, (d) ${\rm Re}\, [S^{(1)}]$,
(e) ${\rm Re}\, [S^{(2)}]$.
The parameters are $k_1=-1, k_2=-2, A_1=1, A_2=2, l_1=3, l_2=4$.}
\label{solitonsolution}
\end{figure}
In figure \ref{solitonsolution}, we see that
waves in $S^{(1)}$ and $S^{(2)}$ have different modulation property,
i.e., carrier waves in $S^{(1)}$ and $S^{(2)}$ has different directions of
propagation. Note that the solutions of equations (\ref{2dls-int}) also
have this property.

It seems that eqs.(\ref{2dls-fluid}) are nonintegrable and
do not admit general
$N$-soliton solution. Similar system (\ref{2dls-int})
has an $N$-soliton solution, but
its physical derivation has not been done yet.

\section{Concluding Remarks}

We have studied solutions of a 
new integrable two-component two-dimensional long
wave-short wave resonant interaction (2c-2d LSRI) system (\ref{2dls-int}). 
We presented a Wronskian formula for 2c-2d LSRI system (\ref{2dls-int}) 
with complex 
conjugacy condition.  
We have also presented solutions of the system (\ref{2dls-fluid})
in the case of two-layer fluid,
i.e. the 2c-2d LSRI system without $Q$. 
In this case, the system (\ref{2dls-fluid}) seems to 
be non-integrable, i.e. the system 
(\ref{2dls-fluid})
does not have multi-soliton solutions. 
We have found that
waves in $S^{(1)}$ and $S^{(2)}$ in both systems
have different modulation property,
i.e., carrier waves in $S^{(1)}$ and $S^{(2)}$ has different directions of
propagation. But the system (\ref{2dls-int}) has much more interesting
solutions such as the V-shape soliton and solitoff 
because of integrability. 

One of authors (K. M.) wishes to acknowledge organizers for providing
the financial support for the ISLAND 3
(Integrable Systems: Linear And Nonlinear Dynamics 3)
conference.


\end{document}